\documentclass[preprint,showpacs,preprintnumbers,amsmath,amssymb]{revtex4}
\usepackage{graphicx}
\usepackage{dcolumn}
\usepackage{bm}

\begin{document}

\title{Topological Excitation in Skyrme Theory}
\author{DUAN Yi-Shi}
\author{ZHANG Xin-Hui}
\thanks{Corresponding author}\email{zhangxingh03@st.lzu.edu.cn}
\author{LIU Yu-Xiao}
\affiliation{Institute of Theoretical Physics, Lanzhou University, Lanzhou 730000, China}

%\date{\today}

\begin{abstract}
Based on the $\phi$-mapping topological current theory and the
decomposition of gauge potential theory, we investigate knotted
vortex lines and monopoles in Skyrme theory and simply discuss the
branch processes (splitting, merging and intersection) during the
evolution of the monopoles.
\end{abstract}


%12.39.Dc Skyrmions
%11.10.Lm Nonlinear or nonlocal theories
%and models (see also 11.27.+d Extended
%classical solutions; cosmic strings, domain walls, texture)
%21.60.Fw Models based on group theory
 \maketitle

\section{Introduction}
The Skyrme theory \cite{a1} has played an important role in
physics, in particular in nuclear physics as a successful
effective field theory of strong interaction. Witten \cite{d5, d6}
revived this idea and proposed that baryons may be described
phenomenologically as the topological solitons, which by now are
called skyrmions. In this view the skyrmions are interpreted as
the baryons and much evidence has been put forward \cite{ a2, a3,
a4}. Meanwhile, it has been discovered that the theory allows not
only the topological skyrmions but also knots and monopoles
\cite{choprl2001}. Similar knots have appeared almost everywhere
recently, in atomic physics in two-component Bose-Einstein
condensates \cite{arx7}, in condensed matter physics in multi-gap
superconductors \cite{twoband} even in plasma physics in coronal
loops \cite{arx10}. Recently Ref. \cite{choprl2001} suggested that
the Skyrme theory is just a theory of self-interacting monopoles.
We can see that a remarkable feature of the Skyrme theory is its
rich topological structure. So one should pay much attention to
the topological
characteristic of the Skyrme theory.\\
\indent In this paper, using the $\phi$-mapping topological current
theory \cite{liuxin13,{liuxin15}} and the decomposition of gauge
potential theory \cite{liuxinprd,{liuxin12}}, we research the inner
topological structure of knots and monopoles in Skyrme theory. If we
write the field $U$ in SU(2) space in a general form
\begin{eqnarray}
U&=&e^{inF(\varphi)}=\cos{F}+in\sin{F}, \label{uu}\\
n&=&n^a\sigma_a,\;\;\;\;\;n^a=\frac{\phi^a}{\|\phi\|},\label{n}
\end{eqnarray}
where $\phi^a\;(a=1,2,3)$ are three fundamental functions with
space-time coordinates $x^\mu$ and $\sigma^a$ are pauli matrices,
the Skyrme model is denoted by the Lagrangian  \cite{choprl2001}
\begin{equation}\label{sl}
L=\frac{u^{2}}{4}\textrm{Tr}
(U^+\partial_{\mu}U)^2+\frac{\alpha}{32}\textrm{Tr}
([U^+\partial_{\mu}U,U^+\partial_{\nu}U])^{2},
\end{equation}
where $u$ and $\alpha$ are the coupling constants. It can be seen
that the Lagrangian has a local U(1) symmetry as well as a global
SU(2) symmetry \cite{cho7}.  Restricting $\varphi$ to $\pi$, one can
reduce Eq. (\ref{sl}) to the Skyrme-Faddeev Lagrangian
\begin{equation}\label{sf}
L_{sf}=-\frac{u^{2}}{2}(\partial_{\mu}\hat{n})^{2}
-\frac{\alpha}{4}(\partial_{\mu}\hat{n}\times\partial_{\nu}\hat{n})^{2},
\end{equation}
whose equation of motion is given by:
\begin{equation}
\hat{n}\times\partial^{2}\hat{n}+\frac{\alpha}{u^{2}}(\partial_{\mu}H_{\mu\nu})\partial_{\nu}\hat{n}=0.
\end{equation}
Here the tensor $H_{\mu\nu}$ is denoted by
\begin{equation}\label{H}
H_{\mu\nu}=\hat{n}\cdot(\partial_{\mu}\hat{n}\times\partial_{\nu}\hat{n})
=\epsilon_{abc}n^a\partial_\mu{n^b}\partial_\nu{n^c},\;\;(a,b,c=1,2,3),
\end{equation}
which is a topological term describing the non-uniform
distribution of the unit vector $\hat{n}$ at large distances in
space. We will explore the topological properties of the Skyrme
theory by analyzing
the tensor $H_{\mu\nu}$.\\
\indent The paper is arranged as follows. In Sec. \ref{sec2},
using the $\phi$-mapping topological current theory and the
decomposition of  $U(1)$ gauge potential theory, we research the
vortex lines in the Skyrme theory and knotted vortex excitation.
The inner topological structure and the evolution of the monopoles
will be studied in Sec. \ref{sec3}. The conclusion of this paper
is given in Sec. \ref{sec4}.

\section{knotted vortex excitation in Skyrme theory}\label{sec2}

In this section, based on the $\phi$-mapping topological current
theory, we show that there is a two-dimensional topological
current, which can be derived from $H_{\mu\nu}$, and the vortex
lines in the skyrme theory are just inhering in this topological
current. One can prove that the tensor $H_{\mu\nu}$ can be
reexpressed as \cite{ho9,WY}
\begin{equation}
H_{\mu\nu}=\partial_\mu{b_\nu}-\partial_\nu{b_\mu},
\end{equation}
where $b_\mu$ is the Wu-Yang potential \cite{WY1}
\begin{equation}\label{bu}
b_\mu=\vec{e}_1\cdot\partial_\mu\vec{e}_2,
\end{equation}
in which $\vec{e_1}$ and $\vec{e_2}$ are two perpendicular unit
vectors normal to $\vec{n}$, and $(\vec{e}_1,\vec{e}_2,\vec{n})$
forms an orthogonal frame:
\begin{equation}\label{ea}
\vec{n}=\vec{e_1}\times{\vec{e_2}},\;\; \vec{e}_1\cdot{\vec{e}_2}=0.
\end{equation} Now, consider a two-component vector field
$\vec{\phi}=(\phi^1,\phi^2)$ residing in the plane formed by
$\vec{e}_1$ and $\vec{e}_2$:
\begin{equation}\label{eaea}
e^a_1=\frac{\phi^a}{\|\phi\|},\;\;e^a_2=\epsilon_{ab}\frac{\phi^b}{\|\phi\|},\;\;
(\|\phi\|^2=\phi^a\phi^a;\; a,b=1,2).
\end{equation}
It can be proved that the expression Eq. (\ref{eaea}) for
$\vec{e}_1$ and $\vec{e}_2$ satisfies the restriction Eq.
(\ref{ea}). Obviously the zero points of $\vec{\phi}$ are just the
singular points of $\vec{e}_1$ and $\vec{e}_2$.
 Using the $\vec\phi$ field, the Wu-Yang potential Eq. (\ref{bu}) can be written
 as
\begin{equation}\label{wy}
b_\mu=\epsilon_{ab}\frac{\phi^a}{\|\phi\|}\partial_{\mu}\frac{\phi^b}{\|\phi\|}.
\end{equation}
Comparing the above formula with the decomposition of $U(1)$ gauge
potential theory \cite{liuxinprd}, we can see that Eq. (\ref{wy}) is
just the expression of the $U(1)$ gauge potential decomposition,
$b_\mu$ satisfies the $U(1)$ gauge transformation. Then the tensor
$H_{\mu\nu}$ is reformed in terms of $\vec\phi(x)$
\begin{equation}
H_{\mu\nu}=
2\epsilon_{ab}\partial_\mu\frac{\phi^a}{\|\phi\|}\partial_\nu\frac{\phi^b}{\|\phi\|}.
\end{equation}
According to Ref. \cite{liuxin15}, using
$\partial_\mu\frac{\phi^a}{\|\phi\|}=\phi^a\partial_\mu\frac{1}{\|\phi\|}+\frac{\partial_\mu\phi}{\|\phi\|}$
and the Green function relation in $\phi$-space:
$\partial_a\partial_aln\|\phi\|=2\pi\delta(\vec\phi)$
($\partial_a=\partial/\partial{\phi^a}$), it can be proved that
\begin{equation}
\frac{1}{4\pi}\epsilon^{\mu\nu\lambda\rho}\epsilon_{ab}
\partial_\lambda\frac{\phi^a}{\|\phi\|}\partial_\rho\frac{\phi^b}{\|\phi\|}
=\delta^2(\vec\phi)D^{\mu\nu}(\frac{\phi}{x}),
\end{equation}
where
$D^{\mu\nu}(\frac{\phi}{x})\!\!=\!\!\frac{1}{2}\epsilon^{\mu\nu\lambda\rho}\epsilon_{ab}
\partial_\lambda\phi^a\partial_\rho\phi^b$.
Indeed if we introduce
$\widetilde{H}^{\mu\nu}\!\!=\!\!\frac{1}{8\pi}\epsilon^{\mu\nu\lambda\rho}H_{\lambda\rho}$,
then we get
$\widetilde{H}^{\mu\nu}=\delta^2(\vec\phi)D^{\mu\nu}(\frac{\phi}{x})$.
Defining the spacial components of $\widetilde{H}^{\mu\nu}$ as
\begin{equation}\label{hi}
H^i=\widetilde{H}^{0i}=\delta^2(\vec\phi)D^{i}(\frac{\phi}{x}),\;\;(i=1,2,3),
\end{equation}
where $D^i(\phi/x)=D^{0i}(\phi/x)$. From Eq. (\ref{hi}), we can
see that $H^{i}$ exists if and only if $\vec{\phi}=0$. So it is
necessary to study the zero points of $\vec{\phi}$ to determine
the nonzero solutions of $H^i$.  The implicit function theory
\cite{liuxin12} shows that under the regular condition
\begin{equation}
D^{i}(\frac{\phi}{x})\neq0,
\end{equation}
the general solutions of $\phi^{a}(\vec{x})=0$ represent the world
surface of $N$ moving isolated singular strings
$L_{k}\;(k=1,2\cdot\cdot\cdot{N})$ with string parameter $s$.
These singular string solutions are just the vortex lines in the
Skyrme theory. In $\delta$-function theory \cite{liuxinprd17}, one
can prove that in three-dimensional space
\begin{equation}\label{delta}
\delta^{2}(\vec{\phi})=\sum^{N}_{k=1}\beta_k\int_{L_{k}}
\frac{\delta^{3}(\vec{x}-\vec{x}_{k}(s))}{|D(\frac{\phi}{u})|_{\Sigma_{k}}}ds,
\end{equation}
where $D(\frac{\phi}{u})=\frac{1}{2}\epsilon^{jk}\epsilon_{mn}
(\partial\phi^{m}/\partial{u}^{j})(\partial\phi^{n}/\partial{u}^{k})$
and $\Sigma_{k}$ is the $k$-th planar element transverse to
$L_{k}$ with local coordinates $(u^{1}, u^{2})$. The positive
integer $\beta_{k}$ is the Hopf index of $\phi$-mapping, which
means that when $\vec{x}$ covers the neighborhood of the zero
point $\vec{x}_{k}(s)$ once, the vector field $\vec{\phi}$  covers
the corresponding region in $\phi$ space $\beta_{k}$ times.
Meanwhile the direction vector of $L_{k}$ is given by
\cite{liuxin15}
\begin{equation}\label{velocity}
\left.\frac{dx^{i}}{ds}\right|_{\vec{x}_{k}}=\left.\frac{D^{i}(\frac{\phi}{x})}{D(\frac{\phi}{u})}\right|_{\vec{x}_{k}}.
\end{equation}
Then considering Eqs. (\ref{delta}) and (\ref{velocity}), we
obtain the inner structure of $H^{i}$
\begin{equation}\label{jcurrent}
H^{i}=\delta^{2}(\vec\phi)D^{i}(\frac{\phi}{x})=\sum^{N}_{k=1}\beta_{k}\eta_k\int_{L_{k}}
\frac{dx^{i}}{ds}\delta^{3}(\vec{x}-\vec{x}_{k}(s))ds,
\end{equation}
where $\eta_{k}=\textrm{sgn}D(\frac{\phi}{u})=\pm1$ is the Brouwer
degree of $\phi$-mapping, with $\eta_{k}=1$ corresponding to the
vortex and $\eta=-1$ corresponding to the antivortex. From Eq.
(\ref{jcurrent}) one can obtain the topological number of vortex
line $L_k$
\begin{equation}
Q_k=\int_{\Sigma_k}H^id\sigma_i=W_k,
\end{equation}
in which $W_{k}=\beta_{k}\eta_{k}$ is the winding number of
$\vec{\phi}$ around $L_{k}$. \\
\indent Now we discuss the topological property of knotted vortex
lines in the Skyrme system. First define a action, which is based
on the so-called Hopf invariant
\begin{equation}\label{action1}
I=\frac{1}{16\pi^2}\int\epsilon^{ijk}{b}_i{H}_{jk}d^3x=\frac{1}{2\pi}\int{b_i}H^id^3x,
\end{equation}
where $i; j; k=1;2;3$ denote the 3-dimensional space. Substituting
Eq. (\ref{jcurrent}) into Eq. (\ref{action1}), we get
\begin{equation}\label{action2}
I=\frac{1}{2\pi}\sum_{k=1}^NW_k\int_{L_k}b_i{dx^i}.
\end{equation}
It can be seen that when these $N$ vortex lines are $N$ closed
curves, i.e., a family of $N$ knots $\gamma_k$
$(k=1,\cdot\cdot\cdot,N),$ then Eq. (\ref{action2}) leads to
\begin{equation}\label{action3}
I=\frac{1}{2\pi}\sum_{k=1}^NW_k\oint_{\gamma_k}b_i{dx^i}.
\end{equation}
This is a very important expression. Consider the $U(1)$ gauge
transformation of $b_\mu$: $b_\mu'=b_\mu+\partial_i\theta$, where
$\theta\in{I\!\!R}$ is a phase factor denoting the $U(1)$
transformation. It is seen that the $\partial_i\theta$ term in Eq.
(\ref{action3}) contributes noting to the integral $I$, hence the
expression (\ref{action3}) is invariant under the $U(1)$ gauge
transformation. Meanwhile we know that $I$ is independent of the
metric $g_{\mu\nu}$ \cite{liuxinprd}. Therefore one can conclude
that $I$ is a topological invariant  for the knotted vortex lines
(a knot invariant) in the Skyrme theory.

\section{monopoles in Skyrme theory}\label{sec3}

In this section, we will explore the inner topological structure
of monopoles in Skyrme theory by using the $\phi$-mapping
topological theory. The generalized winding number $W$ can be
obtained  by integrating  on a closed surface $\partial\Omega$,
where $\Omega$ is a spacial volume and $\partial\Omega$ is its
boundary. $W$ is defined by the Gauss map $n:
\partial\Omega\rightarrow{S^2}$ \cite{ho9}
\begin{equation}
W=\frac{1}{8\pi}\int_{\partial\Omega}n^*(\epsilon_{abc}n^adn^b\wedge{dn}^c),
\end{equation}
where $n^*$ is the pullback of the map $n$. Then we have
\begin{eqnarray}
W&=&\frac{1}{8\pi}\int_{\partial\Omega}\epsilon_{abc}n^a
\partial_i{n^b}\partial_j{n^c}dx^i\wedge{dx^j} \nonumber \\
&=&\frac{1}{8\pi}\int_{\partial\Omega}H_{ij}dx^i\wedge{dx^j}
\;\;\;(i,j=1,2,3).
\end{eqnarray}
In topology this means that when $\vec{x}$ covers $\partial\Omega$
in the real space once, the unit vector $\hat{n}$ will cover $S^2$
$W$ times. This is a topological invariant and is called the
degree of Gauss map. At the same time, $W$ is also the total
topological charge of the point defects (i.e. the singular points
of $\hat{n}$) located in the volume $\Omega$. Using Stokes'
theorem, we have
\begin{equation}
W=\frac{1}{8\pi}\int_\Omega\epsilon^{ijk}\epsilon_{abc}\partial_in^a\partial_j
n^b\partial_kn^cd^3x=\int_\Omega\rho{d^3x},
\end{equation}
where the density of point defects is derived from $H_{ij}$,
\begin{equation}
\rho=\frac{1}{8\pi}\epsilon^{ijk}\epsilon_{abc}\partial_in^a\partial_j
n^b\partial_kn^c=\frac{1}{8\pi}\epsilon^{ijk}\partial_iH_{jk}.
\end{equation}
According to the $\phi$-mapping theory, the topological current is
defined as
\begin{equation}
J^\mu=\frac{1}{8\pi}\epsilon^{\mu\nu\lambda\rho}\epsilon_{abc}
\partial_\nu{n^a}\partial_\lambda{n^b}\partial_\rho{n^c},\;\;(\mu=0,1,2,3),
\end{equation}
obviously, $\rho$ is the $0$-th component of $J^\mu$. Taking
account of Eq. (\ref{n}) and using
$\partial_\mu{n}^a=\phi^a\partial_\mu({1/\|\phi\|})+(1/\|\phi\|)\partial_\mu\phi^a$,
$J^\mu$ can be expressed as
\begin{equation}
J^\mu=\frac{1}{8\pi}\epsilon^{\mu\nu\lambda\rho}\epsilon_{abc}
\frac{\partial}{\partial\phi^l}\frac{\partial}{\partial\phi^a}ln(\|\phi\|)
\partial_{\nu}\phi^l\partial_{\lambda}\phi^b\partial_{\rho}\phi^c.
\end{equation}
If we define Jacobian
$\epsilon^{lbc}D^\mu(\frac{\phi}{x})=\epsilon^{\mu\nu\lambda\rho}\partial_{\nu}\phi^l
\partial_{\lambda}\phi^b\partial_{\rho}\phi^c,$
and make use of the Green function relation in ${\phi}$ space
\cite{liuxin12}, we obtain the $\delta$-function like current
\begin{equation}\label{bu2}
J^\mu=\delta(\vec{\phi})D^\mu(\frac{\phi}{x}).
\end{equation}
The expression of Eq. (\ref{bu2}) provides an important conclusion:
\begin{equation}
J^\mu\left\{
\begin{array}{l}
=0,\;\textrm {if and only if}\;{\vec{\phi}}\neq 0. \\
\neq 0,\;\textrm {if and only if}\;{\vec{\phi}}=0.
\end{array}
\right.
\end{equation}So it is
necessary to study the zero points of $\vec\phi$ to determine the
nonzero solutions of $J^\mu$. The implicit function theory
\cite{liuxin16} shows that under the regular condition
\begin{equation}
D^{\mu}(\frac{\phi}{x})\neq0,
\end{equation}
the general solutions of
\begin{eqnarray}
\phi^{1}(x^{1}, x^{2},x^{3}, t)&=&0, \nonumber \\
\phi^{2}(x^{1}, x^{2},x^{3}, t)&=&0,\\
\phi^{3}(x^{1}, x^{2},x^{3}, t)&=&0    \nonumber
\end{eqnarray}
can be expressed as
\begin{equation}\label{phi}
x^{1}=x^{1}_{k}(t),\;\;\;x^{2}=x^{2}_{k}(t),\;\;\;x^{3}=x^{3}_{k}(t),
\end{equation}which represent the world lines of $N$ moving isolated singular
pointlike solutions. These singular solutions are just the
monopoles in Skyrme theory.\\
\indent Above we have used the regular condition
$D^{\mu}(\frac{\phi}{x})\neq0$. When this condition fails, branch
processes will occur. It is known the velocity of the $k$-th zero
is determined by
\begin{equation}\label{velocity2}
\frac{dx^{i}}{dt}=\left.\frac{D^{i}(\frac{\phi}{x})}{D^{0}{(\frac{\phi}{x})}}
\right|_{x=\vec{x}_{k}},\;\;(i=1,2,3).
\end{equation}
It is obvious that
when $D^{0}(\frac{\phi}{x})=0$, at the very point $({t^{*},
\vec{x}^{*}})$, the velocity
\begin{eqnarray}
\frac{dx^{1}}{dt}&=&\left.\frac{D^{1}(\frac{\phi}{x})}{D^{0}(\frac{\phi}{x})}\right|_{(t^{*},
x^{*})}, \nonumber \\
\frac{dx^{2}}{dt}&=&\left.\frac{D^{2}(\frac{\phi}{x})}{D^{0}(\frac{\phi}{x})}\right|_{(t^{*},
x^{*})}, \\
\frac{dx^{3}}{dt}&=&\left.\frac{D^{3}(\frac{\phi}{x})}{D^{0}(\frac{\phi}{x})}\right|_{(t^{*},
x^{*})},\nonumber
\end{eqnarray}
is not unique in the neighborhood of $(t^{*}, \vec{x}^{*})$. This
very point $(t^{*}, \vec{x}^{*})$ is called the bifurcation point.
Without loss of generality, we discuss only the branch of the
velocity component $(dx^{1}/dt)$ at $(t^{*}, \vec{x}^{*})$. It is
known that the Taylor expansion of the solutions of Eq.
(\ref{phi}) in the neighborhood of $(t^{*}, \vec{x}^{*})$ can
generally be expressed as  \cite{liuxin12, liuxin15}
\begin{equation}
A(x^{1}-x^{1*})^{2}+2B(x^{1}-x^{1*})(t-t^{*})+C(t-t^{*})^{2}+\cdots=0,
\end{equation}
where $A, B$ and $C$ are three constants. Then the above Taylor
expansion leads to
\begin{equation}\label{Taylor}
A\left(\frac{dx^{1}}{dt}\right)^{2}+2B\frac{dx^{1}}{dt}+C=0\;\;\;(A\neq0).
\end{equation}
The solutions of Eq. ({\ref{Taylor}}) give different motion
directions of the zero point at the bifurcation point. There are
two possible cases\\
Case 1.  For $\bigtriangleup=4(B^{2}-AC)=0$, from Eq.
(\ref{Taylor}), we get only one motion direction of the zero point
at the bifurcation point: $(dx^{1}/dt)|_{1, 2}=-B/A$, which includes
three sub-cases: (a) one monopole splits into two monopoles; (b) two
monopoles merge into one monopole; (c) two
monopoles tangentially intersect at the bifurcation point. \\
Case 2. For $\bigtriangleup=4(B^{2}-AC)>0$, from Eq.
(\ref{Taylor}), we get two different motion directions of the zero
point: $(dx^{1}/dt)=(-B\pm\sqrt{B^{2}-AC})/A$. This is the
intersection of two monopoles, which means that the two monopoles
meet and then depart at the bifurcation point.

In both cases 1 and 2, the sum of the topological charges
of final monopole(s) must be equal to that of the initial
monopole(s) at the bifurcation point.

\section{conclusion}\label{sec4}

In this paper, using the decomposition of $U(1)$ gauge potential
theory and the $\phi$-mapping topological current theory, the
inner topological structure of knots and monopoles in Skyrme
theory is researched in detail. In Sec. \ref{sec2}, we revel there
exist topological vortex lines in Faddeev-Skyrme model, mainly
research the knotted vortex lines and obtain a knot invariant. In
Sec. \ref{sec3}, starting with the definition of the winding
number, which is a topological invariant for point defects, we
investigate the inner topological structure of monopoles and
discuss the branch processes (splitting, merging and
intersection). The topological charges are preserved in the branch
processes during the evolution of monopoles.

\section*{Acknowledgment}

This work was supported by the National Natural Science Foundation
and the Doctor Education Fund of Educational Department of the
People's Republic of China.

\end{document}